\documentclass{article}
\usepackage{spconf,amsmath,graphicx}

\usepackage{booktabs}
\usepackage{multirow}
\usepackage{cite}
\usepackage{bm}
\title{emotion controllable speech synthesis on emotion-unlabeled dataset using cross-domain speech emotion recognition}
\title{EMOTION CONTROLLABLE SPEECH SYNTHESIS USING EMOTION-UNLABELED DATASET WITH THE ASSISTANCE OF CROSS-DOMAIN SPEECH EMOTION RECOGNITION}
\name{Xiong Cai$^{1, \dagger}$\thanks{$\dagger$ Equal contribution}, Dongyang Dai$^{1,\dagger}$, Zhiyong Wu$^{1,2, *}$\thanks{* Corresponding author}, Xiang Li$^1$, Jingbei Li$^1$, Helen Meng$^{1,2}$}
\address{
    $^1$Tsinghua-CUHK Joint Research Center for Media Sciences, Technologies and Systems,\\Shenzhen International Graduate School, Tsinghua University, Shenzhen, China \\
    $^2$Department of Systems Engineering and Engineering Management, \\The Chinese University of Hong Kong, Shatin, N.T., Hong Kong SAR, China \\
    \{cai-x18, ddy17, xiang-li20, lijb19\}@mails.tsinghua.edu.cn, \{zywu, hmmeng\}@se.cuhk.edu.hk
}
\address{
    $^1$Shenzhen International Graduate School, Tsinghua University, Shenzhen, China \\
    $^2$The Chinese University of Hong Kong, Shatin, N.T., Hong Kong SAR, China \\
    \{cai-x18, ddy17, xiang-li20, lijb19\}@mails.tsinghua.edu.cn, \{zywu, hmmeng\}@se.cuhk.edu.hk
}

\begin{document}
\ninept
\maketitle
\begin{abstract}
Neural text-to-speech (TTS) approaches generally require a huge number of high quality speech data, which makes it difficult to obtain such a dataset with extra emotion labels. In this paper, we propose a novel approach for emotional TTS synthesis on a TTS dataset without emotion labels. Specifically, our proposed method consists of a cross-domain speech emotion recognition (SER) model and an emotional TTS model. Firstly, we train the cross-domain SER model on both SER and TTS datasets. Then, we use emotion labels on the TTS dataset predicted by the trained SER model to build an auxiliary SER task and jointly train it with the TTS model. Experimental results show that our proposed method can generate speech with the specified emotional expressiveness and nearly no hindering on the speech quality.
\end{abstract}
\vspace{-0.20cm}
\section{Introduction}
\vspace{-0.10cm}
\label{sec1}
In the past few years, neural speech synthesis techniques have experienced great development. End-to-end TTS systems, such as \cite{r01,r02,r03} have achieved remarkable results in terms of naturalness and intelligibility of speech. Benefiting from these techniques, the field of controllable TTS has attracted extensive attention from researchers because it is closer to the practical applications.

The emotion contained in speech is an important paralinguistic information that can effectively expresse the intention of the speaker. Therefore, it is necessary to add emotional control to the TTS system for building a more intelligent interactive interface. 

Many pioneering methods have been proposed for emotional TTS\@. \cite{r04} proposes a LSTM-based acoustic model for emotional TTS, where several kinds of emotional category labels such as one-hot vector or perception vector are used as an extra input to the acoustic model. \cite{r05} uses a improved tacotron\cite{r01} model for end-to-end emotional TTS, in which the emotion labels are concatenated to the output of both the decoder pre-net and the first decoder RNN layer. Some other studies use the global style tokens (GST)\cite{r06} framework to model the emotional features. \cite{r07} proposes an effective style token weights control scheme that uses the centroid of weight vectors of each emotion cluster to generate speech of the emotion. \cite{r08} is also a GST-based method for emotional TTS, where the authors propose an inter-to-intra distance ratio algorithm that well considers the distribution of emotions to determine the emotion weights.

The methods mentioned above report some promising results in the aspect of emotion expressiveness, but these methods rely on an emotion-annotated dataset which is most likely not available. In fact, the lack of emotion-annotated dataset is one of the main obstacles that limit the research of emotional speech synthesis. This problem mainly stems from two reasons: on the one hand, TTS requires a large amount of data, which makes it costly to label emotions; on the other hand, TTS requires high quality of speech data, so dataset from the speech emotion recognition field can not be directly used for emotional speech synthesis. 

Therefore, some semi-supervised approaches have been proposed to alleviate the burden of data requirements. \cite{r09} proposes to fine-tune a pre-trained TTS model on a small emotional dataset for low resource emotional TTS\@. \cite{r10} uses a variant of the GST model and shows that  training with 5\% labeled data can achieve satisfactory results. \cite{r11} proposes to merge an external SER dataset and a labeled subset of  TTS dataset to train a SER model and label the whole TTS dataset by the trained SER model. These semi-supervised methods can greatly reduce the amount of labeled data required for model training. However,  these methods are still not universal enough because a subset of the new dataset still needs to be manually annotated when it is is used for emotional TTS. 

Speech emotion recognition (SER) is another important topic in the field of speech processing. A variety of datasets\cite{r12,r13,r14} are publicly released and a large number of approaches\cite{r15,r16,r17,r19} are proposed in this topic. Therefore, a natural idea is whether we can utilize the achievements in SER to solve the problem of lack of emotion-annotated dataset for emotional TTS\@. In this paper, we propose a novel GST-based model that is trained on a fully emotion-unlabeled dataset and can generate speech with expected emotions. We perform mean opinion score (MOS) evaluations and emotion recognition perception evaluations in 4 \emph{emotion categories} (neutral, happy, sad and angry) and 2 \emph{polarities of emotion dimensions} (high or low for arousal, positive or negative for valence). The evaluation results show that our proposed method can generate speech with the specified emotion expressiveness and nearly no hindering on the speech quality.
\vspace{-0.20cm}
\section{Related work}
\vspace{-0.10cm}
\label{sec2}
The idea most similar to ours is \cite{r20}. The authors propose a general method that enables control of arbitrary variations of speech on a dataset without labels of this variation and validate their idea with an example of emotion control. The authors train a SER model using  an external SER dataset and label the TTS dataset by this trained SER model. Then the predicted labels is fed as an extra input to the statistical parametric speech synthesis (SPSS) model. Different from \cite{r20}, we inject emotion information to the TTS model using an emotion embedding, which can retain more prosody related features rather than a simple emotion label. Meanwhile, as for SER model, we use a domain adaptation technique to reduce distribution shift between SER and TTS datasets, which is also not considered in \cite{r20}. 
\vspace{-0.20cm}
\section{Methodology}
\vspace{-0.1cm}
\label{sec3}
As shown in Figure~\ref{fig:overall_model}, our proposed method includes a cross-domain SER model and a GST--based TTS model. The training and inference procedures of our method are as follows. Firstly, the cross-domain SER model is pre-trained on the emotion labeled SER dataset (as the source domain) and the emotion-unlabeled TTS dataset (as the target domain). Then, the soft emotion labels of the TTS dataset are obtained from the softmax output of the trained SER model (the green dashed arrow). Finally, the TTS model and an emotion predictor are jointly trained on the TTS dataset with the emotion labels. For the inference, we firstly select a \emph{reference audio set} for each emotion category from the TTS dataset. Then, we average the style tokens’ weights of all audios in the reference audio set to synthesize speech for this kind of emotion. 
\vspace{-0.2cm}
\subsection{Cross-domain SER model}
\vspace{-0.1cm}
\label{sec3:sub1}
\subsubsection{Model structure}
\label{sec3:sub1:subsub1}
\begin{figure}[t]
    \scalebox{0.98}{
  \centering
    \includegraphics[width=\linewidth]{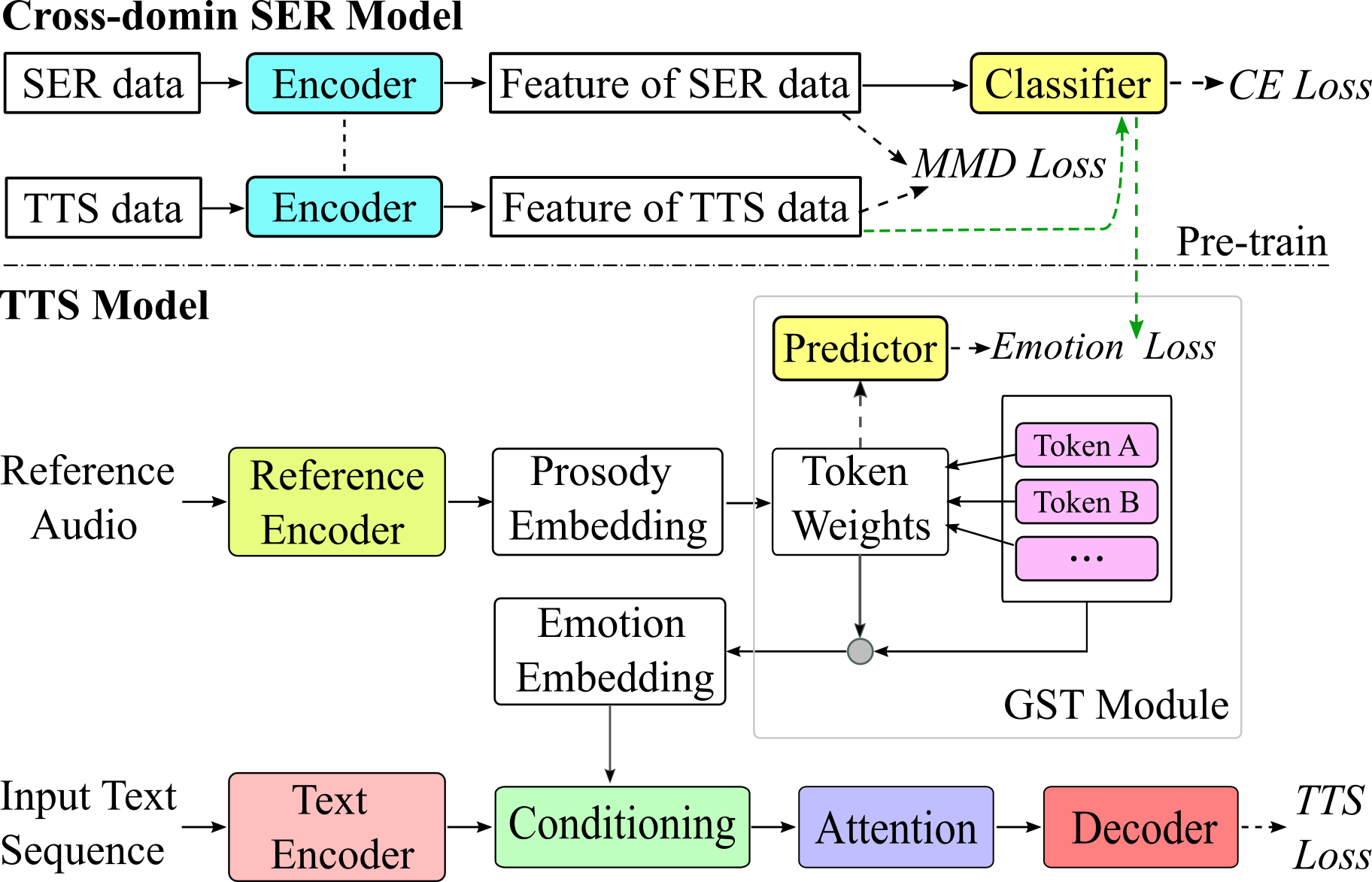}
}
\caption{The overall structure of the cross-domain SER and GST-based TTS model.}
\label{fig:overall_model}
\end{figure}

\vspace{-0.1cm}
As shown in the upper part of Figure~\ref{fig:overall_model}, the SER model consists of a feature extraction encoder and an emotion classifier. The encoder is a CNN-RNN network, which is a popular structure  in SER\@. Firstly, a 4-layer of 2D convolution network takes the 80-dimension log mel spectrum as input feature and outputs a 2D feature map which is flattened into a sequence of feature vectors. Then, a bidirectional GRU layer is followed and outputs the hidden state vector of the last time step as the final feature vector. The emotion classifier is a 2-layer dense network with a softmax activation for the output. The number of output units is 4 for the emotion category classification task and 2 for the arousal and valence polarity  classification task.
\vspace{-0.2cm}
\subsubsection{Cross-domain training}
\vspace{-0.1cm}
\label{sec3:sub1:subsub2}
Since the TTS and SER datasets are quite different in speakers, recording devices and recording environment, some domain adaptation technique is necessary for our SER task to reduce the distribution shift of these two datasets. The Maximum Mean Discrepancy (MMD)\cite{r21} is a kernel-based test statistic to judge whether two distributions are equal, which is widely used in domain adaptation \cite{r22,r23} for measuring the similarity of two distributions. It has also been proved to be effective for cross-domain SER in \cite{r16,r17}. Considering that the training of MMD is stable and our TTS dataset has no available emotion labels for parameter tuning,we choose the MMD method for our cross-domain SER task.

We refer to the SER dataset as the source domain $D_s$ and the TTS dataset as the target domain $D_t$. As described in [23], we minimize the following MMD loss to reduce the distribution discrepancy between the two domains.
\begin{align}
    L_{MMD}=\frac{1}{m^2}\Sigma_{i=1}^{m} &\Sigma_{j=1}^{m}k(\bm{s}_i,\bm{s}_j) + \frac{1}{n^2}\Sigma_{i=1}^{n}\Sigma_{j=1}^{n}k(\bm{t}_i,\bm{t}_j) \notag
                                       \\ & - \frac{1}{mn}\Sigma_{i=1}^{m}\Sigma_{j=1}^{n}k(\bm{s}_i,\bm{t}_j)
\end{align}
Where $\bm{s}_i$ and $\bm{t}_i$ are the output features of the encoder from $D_s$ and $D_t$ respectively; $m$ and $n$  are the number of samples of $D_s$ and $D_t$ respectively; $k(.,.)$ is the kernel function that is a linear combination of multiple RBF kernels:$k(\bm{x}_i, \bm{x}_j) = \Sigma_n\eta_n\exp\{-\frac{1}{2\sigma_n}\|\bm{x}_i-\bm{x}_j\|^2\}$, where $\sigma_n $ is the standard deviation and $\eta_n$ is the weight for the $n$-th RBF kernel. As for the samples from $D_s$, we use a weighted cross-entropy loss for training the emotion classification task.
\begin{align}
L_{CE} = -\Sigma_{i=1}^{m}{\omega_{argmax(\bm{y}_i)}\cdot \bm{y}_i^T\cdot \log(\hat{\bm{y}_i})}
\end{align}
Where $\omega_k$  is the weight of the $k$-th emotion category which is inversely proportional to the number of samples in this emotion category. Therefore, the final loss function for training SER model is:
\begin{align}
    L = L_{CE} + \lambda L_{MMD}
\end{align}
Where $\lambda$ is the weight of MMD loss.
\vspace{-0.3cm}
\subsection{GST--based emotional TTS model}
\vspace{-0.10cm}
\label{sec3:sub2}
\subsubsection{Model structure}
\label{sec3:sub2:subsub1}
\vspace{-0.10cm}
As shown in the lower part of Figure~\ref{fig:overall_model}, our emotional TTS model consists of a TTS module, reference encoder and GST module. The TTS module is a standard Tacotron2 model except that we add a post-net as in \cite{r01} to convert the mel  spectrum to linear spectrum. The reference encoder and GST module are also the same as in \cite{r06} except that we add an auxiliary emotion prediction task to explicitly guide the style tokens to learn emotion-related features. For the generation of waveform, we simply use the Griffin-Lim algorithm\cite{r24} to convert the  predicted linear spectrum to waveform samples, because our goal is mainly to verify the effectiveness of the proposed method rather than to generate high fidelity speech.
\vspace{-0.20cm}
\subsubsection{Emotion prediction task}
\vspace{-0.10cm}
\label{sec3:sub2:subsub2}
The original GST\cite{r06} is designed to unsupervisedly learn the styles from a reference audio, where the style learned by each token is random and uncertain, and therefore uncontrollable. In our emotional TTS task, in order to force the GST module pay more attention to learn the emotion-related styles, we explicitly add an emotion prediction task based on the style token weights, which will also been verified in our experiments to be critical for the emotional expressiveness of speech. 

In this study, we conduct two emotion control methods corresponding to the two most commonly used emotion descriptions: basic emotion categories and emotion dimensions. For the emotion category, the emotion prediction task is a classifier of a single dense layer that takes the style token weight vector as input and outputs a vector of length 4 for the 4 emotion categories (neutral, happy, sad and angry). For the emotion dimensions, since the used SER dataset is originally annotated as discrete scores for emotion dimensions, we build two binary classifiers for predicting arousal and valence polarity respectively as the emotion prediction task. 
Meanwhile, in order to independently control these two orthogonal emotion dimensions, we split the style token weight vector into two valves, which are then fed into the arousal and valence classifiers respectively. 
\vspace{-0.30cm}
\subsubsection{Choice for reference audio set of each emotion class}
\vspace{-0.10cm}
\label{sec3:sub2:subsub3}
There are two common methods for the synthesis of GST model: selecting a reference audio with the desired style or manually specifying the weights of the style tokens. In our experiments, the quality and stability of speech synthesized from a single reference audio heavily depend on the choice of reference audio and the content of the text to be synthesized, which may be due to the fact that the GST module still does not completely disentangle the audio style and text content. Therefore, we manually specify the weight of style tokens by averaging the style token weights of an audio set that belongs to a certain kind of emotion. The audio set can be directly specified as all the utterances with the same emotion label, as used in \cite{r07}, when the TTS dataset has emotion labels. However, our TTS dataset has no ground-truth emotion labels, and only the soft labels predicted from the cross-domain SER model are available. Because the cross-domain SER model is far less reliable than humans, these predicted emotion labels may contain a great number of mispredictions and we can not use all the utterances with the same soft label as the reference audio set. In order to choose a more reliable reference audio set, we propose that only the \emph{K} utterances with highest posterior are selected as the reference audio set, rather than the full set of a certain emotion. Choosing these \emph{K} high-confidence audios can greatly reduce the impact of the prediction error of the SER model and it is a very important choice for generating an emotional speech in our experiments. We set \emph{K}=50 for all of our experiments.
%
\vspace{-0.20cm}
\section{Experiments}
\vspace{-0.20cm}
\label{sec4}
\subsection{Datasets}
\vspace{-0.10cm}
\label{sec4:sub1}
We use the IEMOCAP as the SER dataset and the English dataset of Blizzard Challenge 2013 (BC2013-English) as the TTS dataset for our emotional TTS experiments. In addition, because there are no emotion labels on BC2013-English, we use the an additional SER dataset: RECOLA to verify the effectiveness of MMD on cross-domain SER tasks.

\noindent \textbf{IEMOCAP}\cite{r12} is an audiovisual database of English dyadic conversations performed by ten professional actors, which contains about 12.5 hours and a total of 10,039 utterances. For emotion category schedule, we use the subset of the data that contains only the neutral, angry, sad, happy, excited utterances and merge the excited to the happy category. For arousal and valence dimension schedule, we use all utterances and map the original 5-point label into binary one-hot label with 2.5 as the dividing point.

\noindent \textbf{RECOLA}\cite{r13} is a multimodal database of French dyadic conversations, where continuous arousal and valence label in the range [-1, 1] are annotated at frame level. In our study, we use all the 1,308 freely available utterances from 23 speakers and map the utterance-level average of the original labels to the binary one-hot label with 0.0 as the dividing point.

\noindent \textbf{BC2013-English}\cite{r25} is a audiobook dataset provided by The Voice Factory, where approximately 300 hours of chapter-sized mp3 files and 19 hours of wav files are available. We use a segmented subset of approximately 198 hours data and filter out utterances that are longer than 14 seconds or have more more than 100 characters from this subset. The final dataset used in our experiments is about 73 hours of total 95k utterances. In addition, the stories are read expressively by a single female speaker and this expressive styles is not annotated, which makes it appropriate for our task.
\vspace{-0.30cm}
\subsection{Training setup}
\vspace{-0.10cm}
\label{sec4:sub2}
Our training process involves first training a cross-domain SER model, followed by training an end-to-end GST-based TTS model.

To train the SER model, we use the cross-entropy loss and an additional MMD loss with a batch size of 96. The Adam\cite{r26} optimizer and learning rate schedule: initialized to 3e-5 and fixed to 3e-4 after 100 steps are used to optimize model parameters. The 80-band mel-scale spectrum is extracted frame-wise as input feature to be consistent with the input of TTS model. We randomly select \emph{N} utterances from the source domain dataset as validation set for early stopping (\emph{N}=500 for IEMOCAP and 200 for RECOLA).

To train the emotional TTS model, we use the mean absolute error for the reconstruction of  both mel and linear spectrum and a cross-entropy loss for the auxiliary emotion prediction task with a batch size of 32. We train all the TTS models with 150k steps using the Adam optimizer with a warm-up learning rate schedule: initial rate 2e-3, warm-up steps 4,000 and decay power -0.5. We also set the softmax temperature of attention mechanism to 2.0 in the synthesis phase, which can effectively improve the stability of alignments especially for synthesizing highly expressive emotion speech. We publicly release the audio demos\footnote{https://thuhcsi.github.io/icassp2021-emotion-tts} and all the source codes\footnote{https://github.com/thuhcsi/icassp2021-emotion-tts} online, where more training and model details can be found.
\vspace{-0.55cm}
\begin{table}[th]
  \caption{WA and UA results for baseline and mmd-based models.}
  \label{tab:1}
  \centering
  \scalebox{0.86}{
  \begin{tabular}{ccccccc}
    \toprule
    & & \multicolumn{2}{c}{\textbf{Iem2Rec}} & \multicolumn{2}{c}{\textbf{Rec2Iem}} & \\
    \cmidrule(lr){3-4} \cmidrule(lr){5-6}
    & \textbf{model} & \textbf{arousal} &\textbf{valence} & \textbf{arousal} & \textbf{valence} & \textbf{average} \\
    \midrule
    \multirow{2}{*}{\textbf{WA}} & \textbf{ser-base} & $0.534$ & $0.632$ & $0.617$ & $0.488$ & $0.568$ \\
    & \textbf{ser-mmd}  & $0.538$ & $\textbf{0.720}$ & $\textbf{0.642}$ &  $\textbf{0.500}$ & $\textbf{0.600}$ \\
    \cmidrule(lr){2-7}
    \multirow{2}{*}{\textbf{UA}} & \textbf{ser-base} & $0.529$ & $0.470$ & $0.500$ & $0.505$ & $0.501$ \\
    & \textbf{ser-mmd}  & $\textbf{0.550}$ & $\textbf{0.489}$ & $\textbf{0.543}$ &  $\textbf{0.518}$ & $\textbf{0.525}$ \\
    \bottomrule
  \end{tabular}}
\end{table}
\vspace{-0.6cm}
\subsection{Cross-domain SER results}
\vspace{-0.10cm}
\label{sec4:sub3}
In order to evaluate the effectiveness of MMD for cross-domain SER tasks, we conduct experiments for the arousal and valence emotion dimensions on the IEMOCAP and RECOLA datasets. It is worth noting that the RECOLA dataset is not annotated with emotion category labels, and thus we do not preform experiments for the emotion category task. We train two models: \textbf{ser-base} and \textbf{ser-mmd}, and the structure of the two models is the same as described in section~\ref{sec3:sub1:subsub1}, except that the \textbf{ser-mmd} model has an additional MMD  loss with weight $\lambda$=0.5. We use \textbf{Rec} and \textbf{Iem} to represent the RECOLA and IEMOCAP dataset, and \textbf{Rec2Iem} means training on \textbf{Rec} and testing on \textbf{Iem} and vice versa. Both the Weighted Accuracy (WA) and Unweighted Accuracy (UA) are selected as our evaluation criteria since an imbalance of emotion categories exists on these two datasets. Table~\ref{tab:1} reports the WA and UA results for the models.

As shown in Table~\ref{tab:1}, except for  the experiment of \textbf{Iem2Rec} of arousal, the \textbf{ser-mmd} model shows an improvement of at least 1.2\% compared to the \textbf{ser-base} model on both WA and UA. And the total average improvement of WA and UA over all four experiments is 3.2\% and 2.4\% respectively. This shows that the performance of cross-domain SER can be improved after the discrepancy of feature distribution between two domains is reduced. 
\vspace{-0.20cm}
\subsection{Emotional TTS results}
\vspace{-0.10cm}
\label{sec4:sub4}
The overall quality and emotional expressiveness of synthesized speech are the two most significant evaluation criteria for an emotional TTS system. We perform two subjective experiments for these two criteria using three systems: our proposed system for 4 emotion categories (\textbf{our-4cls}), our proposed system for 2 emotion dimensions (\textbf{our-2d}) and a baseline system (\textbf{base-4cls}) which is the same as \textbf{our-4cls} except that the auxiliary emotion prediction task is not used in training phase. We randomly choose 10 sentences outside the TTS dataset as inference texts, and 20 university students are invited to participate in our subjective experiments.
\subsubsection{Evaluation for the overall quality of speech}
\vspace{-0.10cm}
\label{sec4:sub4:subsub1}
In this section, we perform mean opinion score (MOS) evaluation experiment in terms of the overall quality of speech (naturalness, intelligibility and speech quality).  Table~\ref{tab:2} and Table~\ref{tab:3} report the MOS results of the above three systems. The average score of \textbf{base-4cls} is higher than that of the \textbf{our-4cls} and \textbf{our-2d} models, but the p-values are much greater than the singificance leval of $\alpha=0.05$ indicating  that there is no significant difference between these models. This result suggests that our proposed \textbf{our-4cls} and \textbf{our-2d} models can almost achieve as good speech quality as the baseline system. In addition, it can also be found that the MOS score of happy is greatly lower than the others for both \textbf{base-4cls} and \textbf{our-4cls}. One possible reason is that the SER model itself has lower accuracy for happy in our experiments, which will lead to great differences in the prosody consistency of the top \emph{K} reference audios selected by the SER model. This also indicates that the performance of the cross-domain SER model is critical for our proposed approach. 
\vspace{-0.50cm}
\begin{table}[th]
  \caption{MOS of \textbf{base-4cls} and \textbf{our-4cls} for 4 emotion categories.}
  \label{tab:2}
  \centering
  \scalebox{0.982}{
  \begin{tabular}{ccccccc}
    \toprule
   \textbf{model} & \textbf{neu} & \textbf{ang} & \textbf{hap}  &\textbf{sad} & \textbf{average }& \textbf{p-value}\\
    \midrule
   \textbf{base-4cls} & $3.90$ & $3.84$ & $3.45$ &  $3.74$ & $3.73$ & $-$\\
    \textbf{our-4cls} & $4.12$ & $3.80$ & $3.11$ &  $3.61$ & $3.66$ & $\textbf{0.20}$\\
    \bottomrule
  \end{tabular}}
\end{table}
\vspace{-0.8cm}
\begin{table}[th]
    \caption{MOS of \textbf{our-2d} for arousal and valence dimensions.}
  \label{tab:3}
  \centering
  \scalebox{1.01}{
  \begin{tabular}{ccccccc}
    \toprule
  \textbf{model} & \textbf{low} & \textbf{high} & \textbf{neg} &\textbf{pos} & \textbf{average} & \textbf{p-value}\\
    \midrule
  \textbf{our-2d} & $3.99$ & $3.33$ & $3.91$ &  $3.41$ & $3.66$ & $\textbf{0.18}$\\
    \bottomrule
  \end{tabular}}
\end{table}
\vspace{-0.5cm}
\subsubsection{Evaluation for the emotional expressiveness of speech}
\begin{figure}[htb]
\vspace{-0.3cm}
\begin{minipage}[b]{0.328\linewidth}
  \centering
  \centerline{\includegraphics[width=3.70cm]{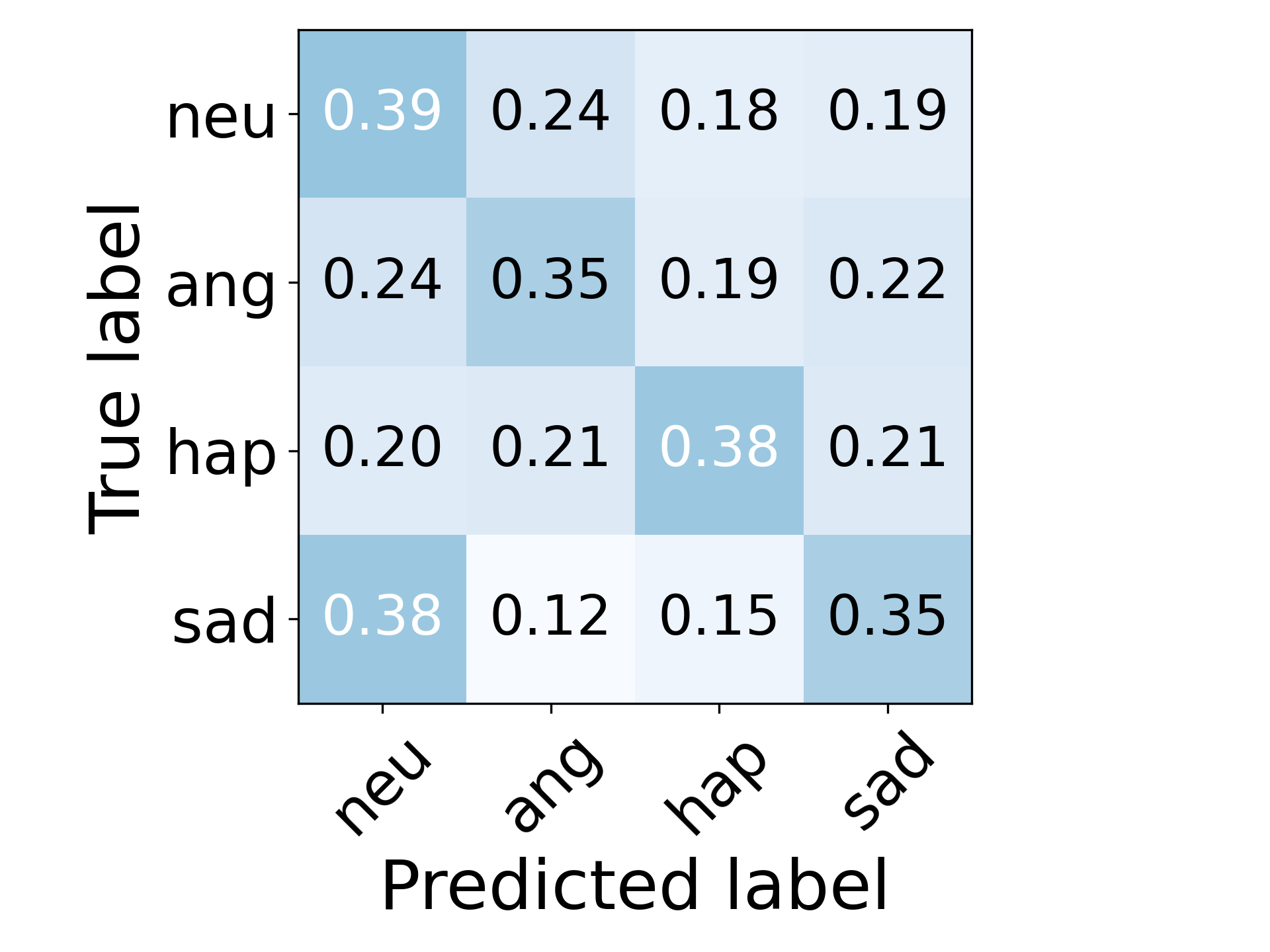}}
  \label{fig:cm_4cls_a}
  \centerline{(a) \textbf{base-4cls}}\medskip
\end{minipage}
\begin{minipage}[b]{0.328\linewidth}
  \centering
  \centerline{\includegraphics[width=3.70cm]{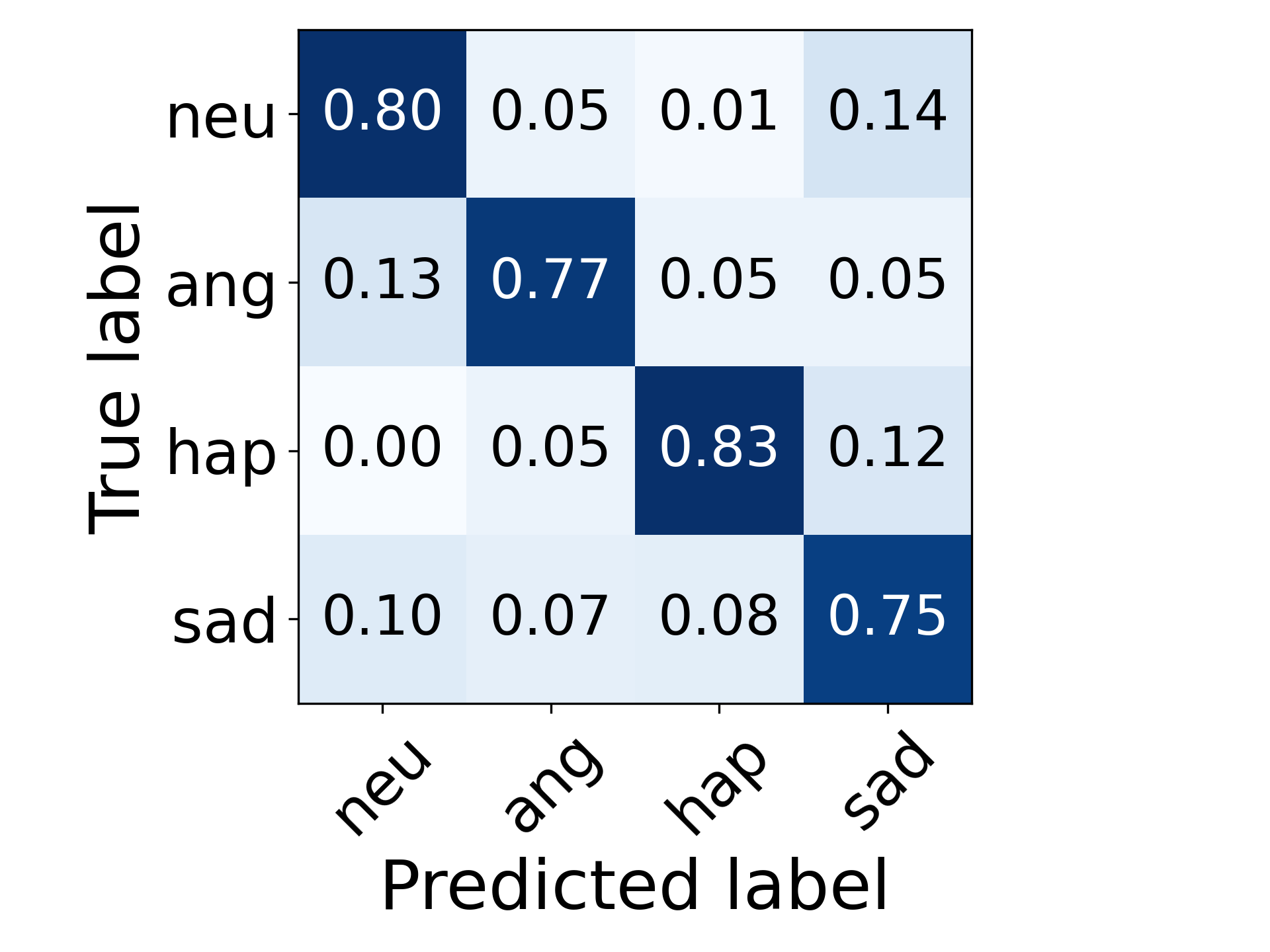}}
  \centerline{(b) \textbf{our-4cls}}\medskip
  \label{fig:cm_4cls_b}
\end{minipage}
\begin{minipage}[b]{0.328\linewidth}
  \centering
  \centerline{\includegraphics[width=3.70cm]{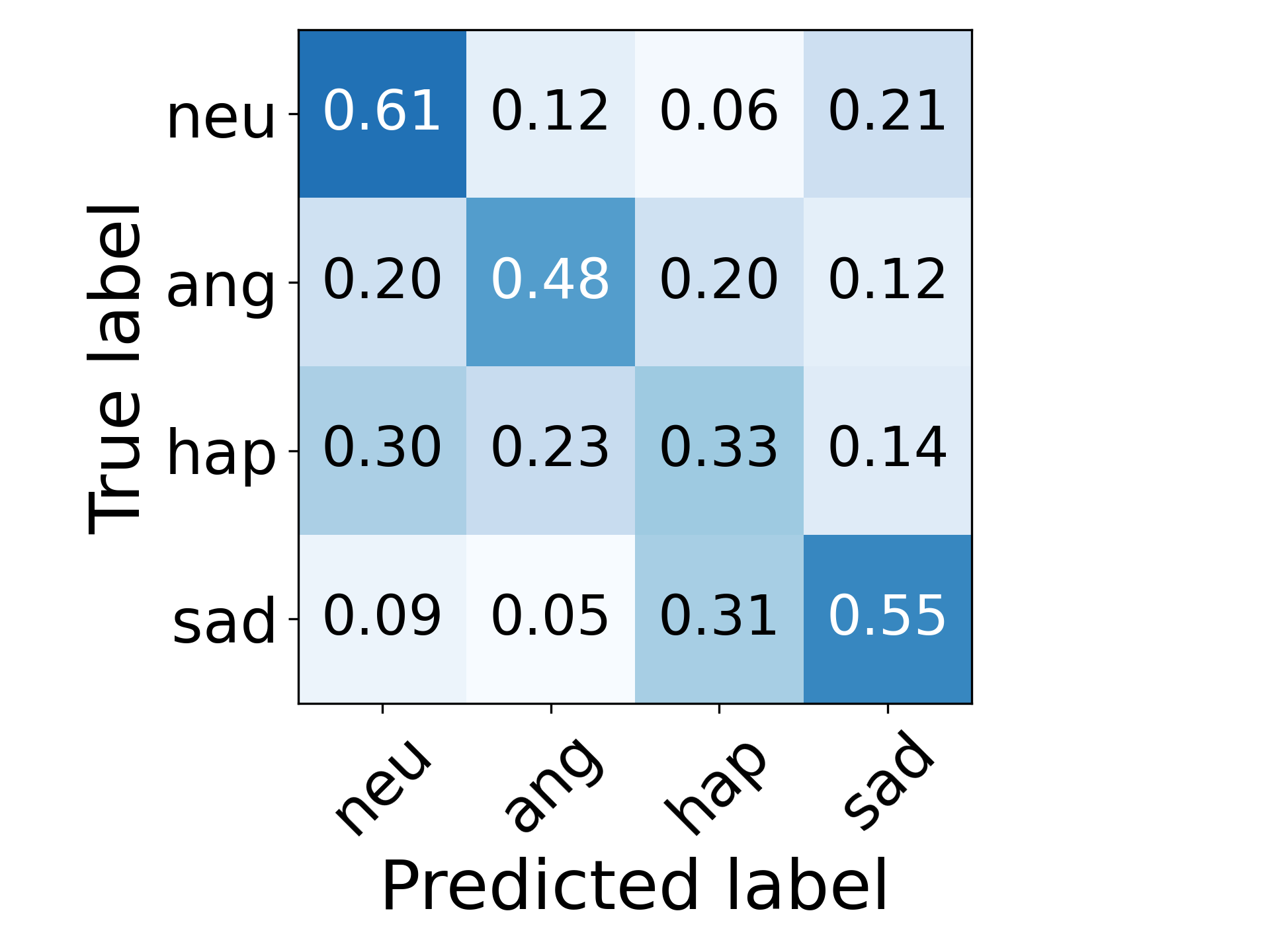}}
  \centerline{(c) \textbf{full-4cls}}\medskip
  \label{fig:cm_4cls_c}
\end{minipage}
\vspace{-0.8cm}
\caption{Confusion matrices of 4 emotion categories for the three methods: \textbf{base-4cls},  \textbf{our-4cls} and \textbf{full-4cls}.}
\label{fig:cm_4cls}
\end{figure}

\begin{figure}[htb]
\vspace{-0.6cm}
\begin{minipage}[b]{0.48\linewidth}
  \centering
  \centerline{\includegraphics[width=3.70cm]{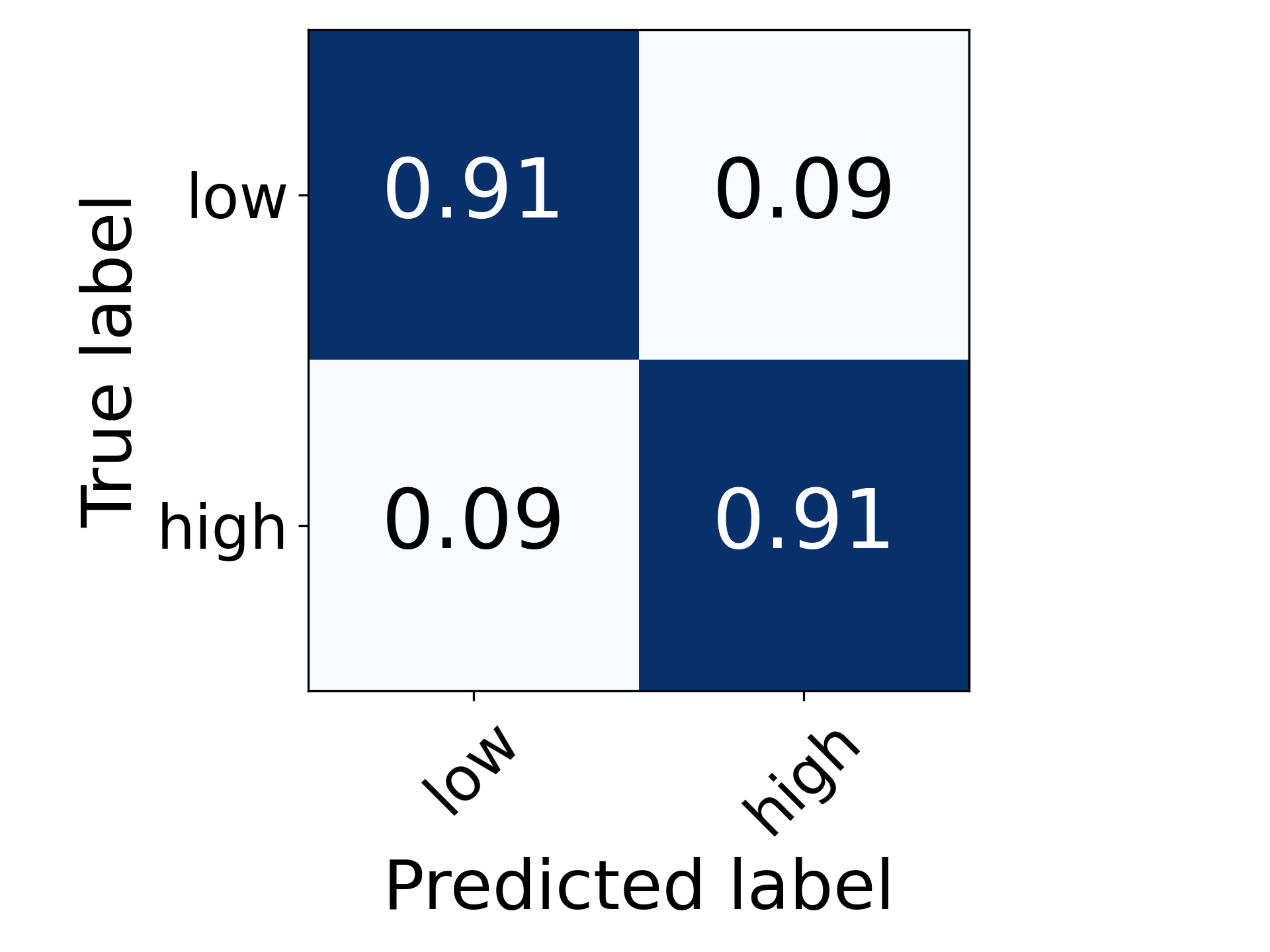}}
  \centerline{(a) arousal dimension}\medskip
  \label{fig:cm_2d_a}
\end{minipage}
\begin{minipage}[b]{0.48\linewidth}
  \centering
  \centerline{\includegraphics[width=3.70cm]{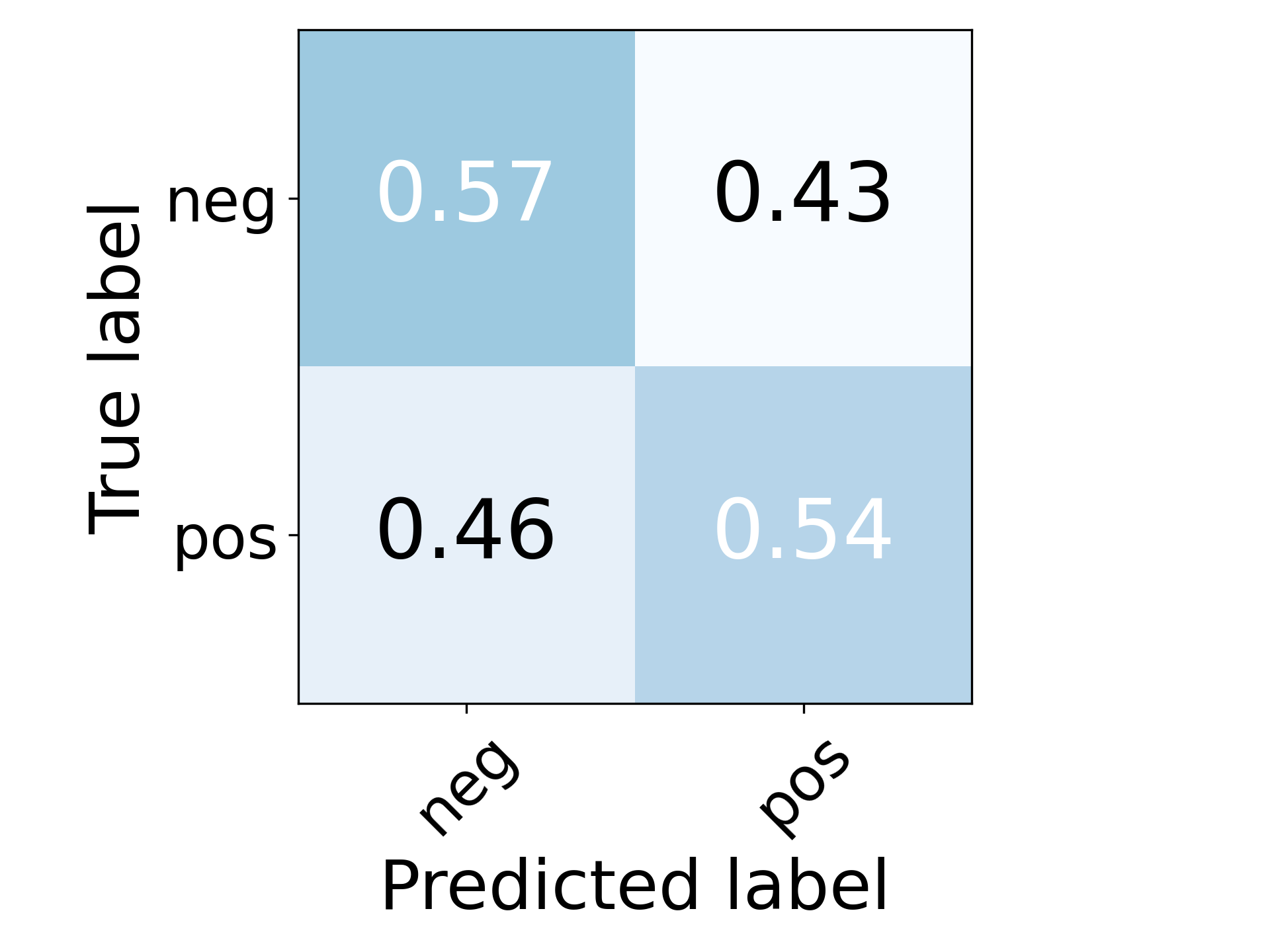}}
  \centerline{(b) valence dimension }\medskip
  \label{fig:cm_2d_b}
\end{minipage}
 \vspace{-0.5cm}
\caption{Confusion matrices of polarities of arousal and valence emotion dimensions  for the model: \textbf{our-2d}.}
\label{fig:cm_2d}
\end{figure}

\vspace{-0.10cm}
\label{sec4:sub4:subsub2}
In this section, we carry out emotion perception experiments in terms of emotional expressiveness. Specifically, for a given text, we generate a set of audios for all emotion categories and then randomly shuffle these audios. Then, the subjects are asked to choose a most likely emotion category for each audio, according to two reference audios given for each emotion. In order to verify the role of our proposed top-\emph{K} scheme, we also add a scheme \textbf{full-4cls} that use the same checkpoint with \textbf{our-4cls} but chooses the full set of audios predicted by the SER model as the same emotion category as the reference audio set. Figure~\ref{fig:cm_4cls} and Figure~\ref{fig:cm_2d} show the confusion matrices of the subjective emotion prediction results. 

We can find that the average accuracies of both \textbf{our-4cls} (78.75\%) and \textbf{full-4cls} (49.25\%) are much higher than the \textbf{base-4cls}’s 36.75\%. This result shows that explicitly adding an emotion prediction task based on style token weights can help the GST module to more effectively the extract emotion-related feature. Moreover, a large gap in terms of average accuracy can also be found between \textbf{our-4cls} and \textbf{full-4cls}, which proves that the proposed top-\emph{K} scheme is fairly effective for the emotional expressiveness. Similarly, as shown in Figure~\ref{fig:cm_2d}, the average accuracy of arousal and valence is 91.0\% and 55.5\% respectively, which are greater than the random level of 50.0\%. This also suggests that our proposed systems can effectively model the emotional expressiveness in speech.

\begin{figure}[htb]
\vspace{-0.6cm}
\begin{minipage}[b]{0.48\linewidth}
  \centering
  \centerline{\includegraphics[width=3.5cm]{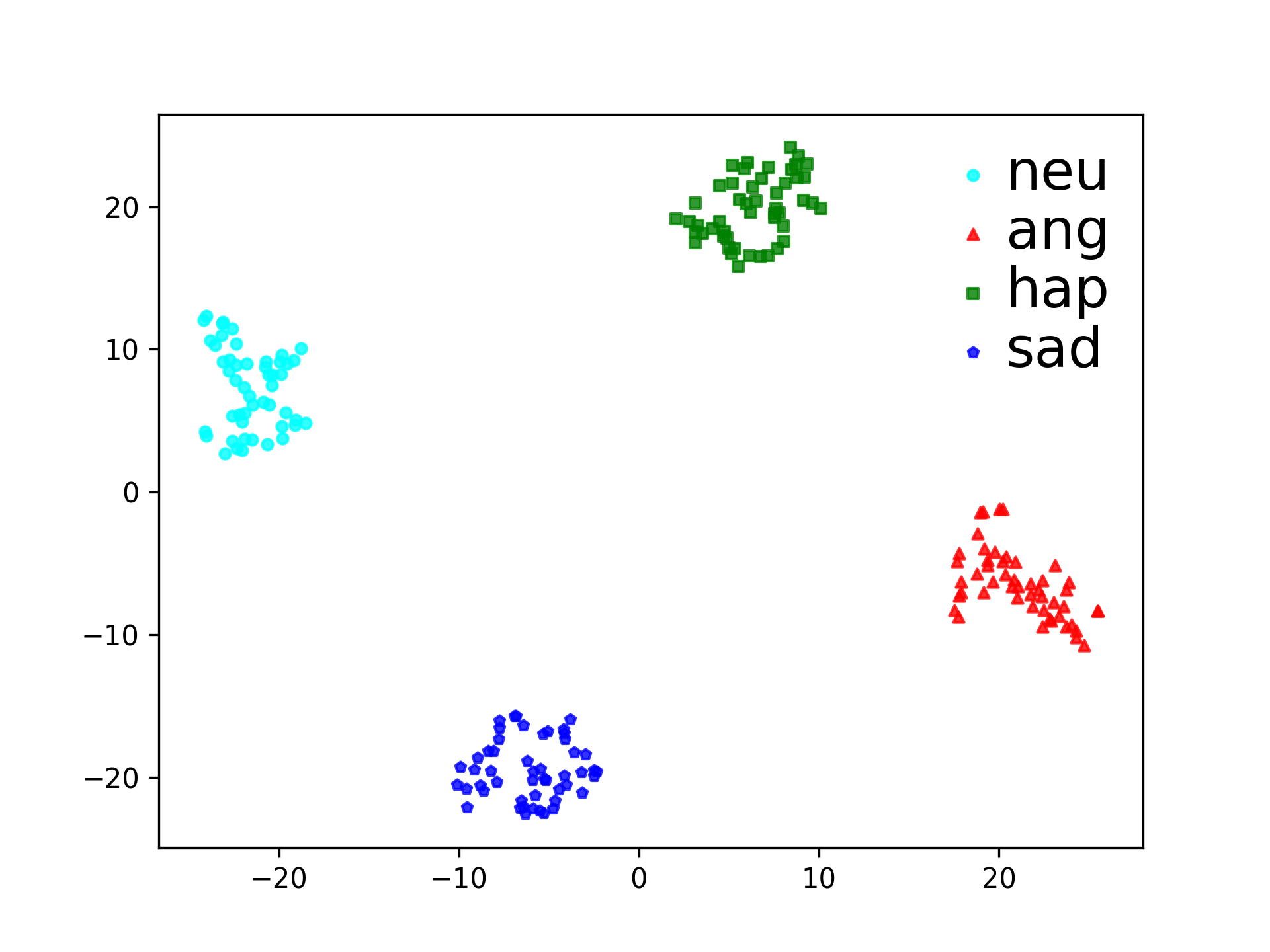}}
  \centerline{(a) \textbf{our-4cls}}\medskip
  \label{fig:visual_a}
\end{minipage}
\begin{minipage}[b]{0.48\linewidth}
  \centering
  \centerline{\includegraphics[width=3.5cm]{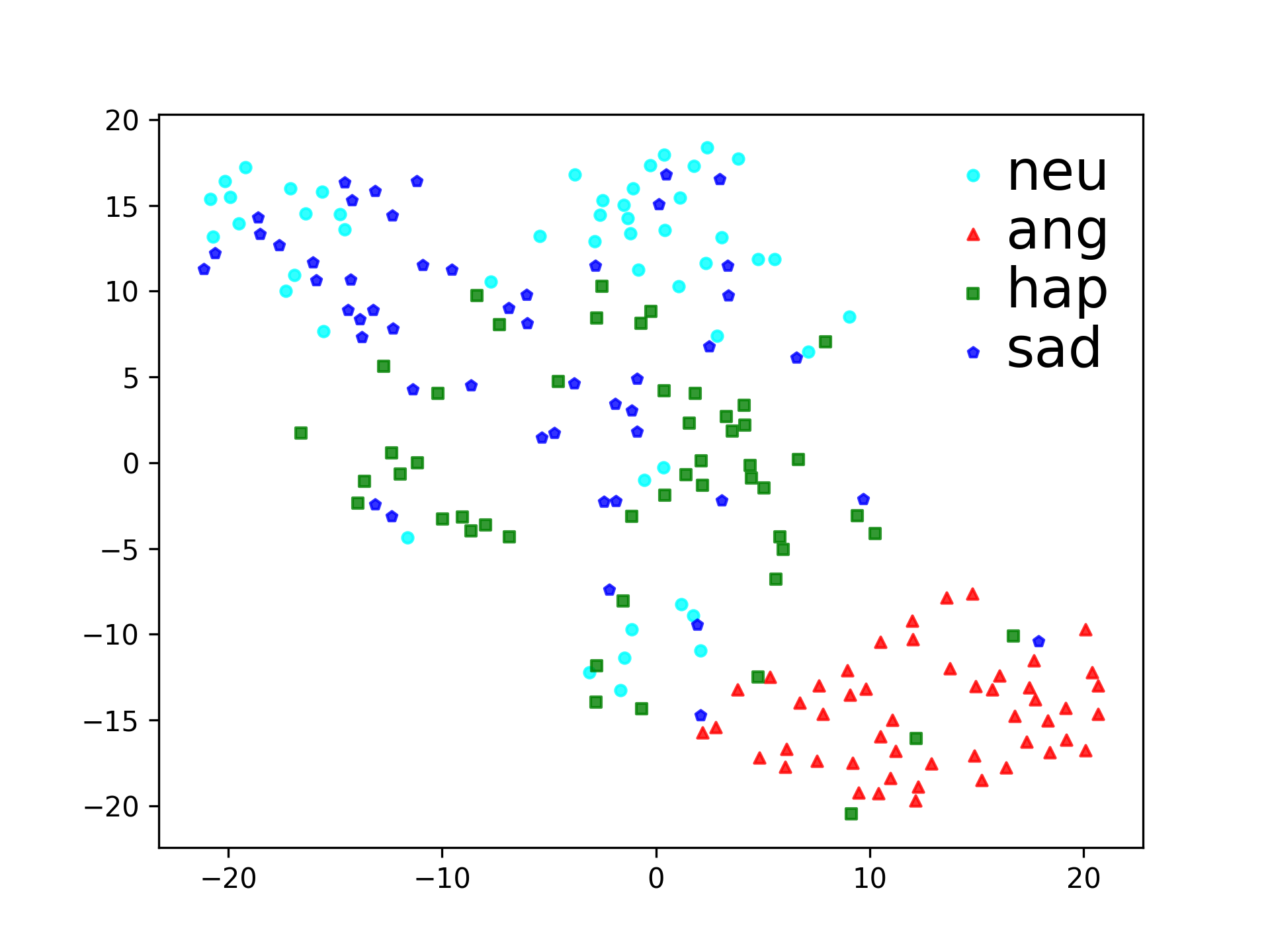}}
  \centerline{(b) \textbf{base-4cls} }\medskip
  \label{fig:visual_b}
\end{minipage}
 \vspace{-0.5cm}
 \caption{The t-SNE 2D visualization for style token weights on the reference audio set for the  \textbf{our-4cls} and \textbf{base-4cls} model.}
\label{fig:visual}
\end{figure}

Finally, we also use the t-SNE\cite{r28} algorithm to project the style token weights on the reference audio sets into the 2D space for \textbf{our-4cls} and \textbf{base-4cls}. As shown in Figure~\ref{fig:visual}, the style token weights of the 4 emotion categories are clearly clustered into 4 clusters for the \textbf{our-4cls} model, while for the \textbf{base-4cls} model, except for the angry catetory, there is no obvious clustering boundary for the other three categories. This can also be considered as another possible evidience to explain the role of the auxiliary emotion prediction task.
\vspace{-0.70cm}
\section{Conclusions}
\vspace{-0.10cm}
\label{sec5}
In this paper, we propose a novel GST-based approach for emotional speech synthesis. Our proposed approach has the following three characteristics: 1) only requires an expressive but emotion-unlabeled TTS dataset; 2) can generate speech with a desired emotional expressiveness; 3) nearly do not hurt the quality of synthesized speech, except for some highly expressive utterances. And three key technologies ensure that our proposed approach works well: 1) an MMD-based cross-domain SER model provides effective emotion labels for the TTS dataset; 2) an auxiliary supervised  emotion prediction task based on the weight of style tokens guides the GST module to model the emotion-related feature more thoroughly; 3) the top-\emph{K} scheme is used to choose the reference audio set for each emotion category. Our proposed approach greatly reduces the threshold of emotional speech synthesis in terms of emotion-annotated data, and its main idea can also be easily applied to other neural TTS systems. For future works, we will explore better decoupling in the control of arousal and valence to achieve more flexible emotional control.


\bibliographystyle{IEEEbib}
\bibliography{refbib}

\begin{thebibliography}{10}

\bibitem{r01}
Yuxuan Wang, RJ~Skerry-Ryan, Daisy Stanton, Yonghui Wu, Ron~J Weiss, Navdeep
  Jaitly, Zongheng Yang, Ying Xiao, Zhifeng Chen, Samy Bengio, et~al.,
\newblock ``Tacotron: Towards end-to-end speech synthesis,''
\newblock in {\em Conference of the International Speech Communication
  Association (INTERSPEECH)}, 2017.

\bibitem{r02}
Jonathan Shen, Ruoming Pang, Ron~J Weiss, Mike Schuster, Navdeep Jaitly,
  Zongheng Yang, Zhifeng Chen, Yu~Zhang, Yuxuan Wang, Rj~Skerrv-Ryan, et~al.,
\newblock ``Natural tts synthesis by conditioning wavenet on mel spectrogram
  predictions,''
\newblock in {\em IEEE International Conference on Acoustics, Speech and Signal
  Processing (ICASSP)}, 2018, pp. 4779--4783.

\bibitem{r03}
Naihan Li, Shujie Liu, Yanqing Liu, Sheng Zhao, and Ming Liu,
\newblock ``Neural speech synthesis with transformer network,''
\newblock in {\em Proceedings of the AAAI Conference on Artificial Intelligence
  (AAAI)}, 2019, vol.~33, pp. 6706--6713.

\bibitem{r04}
Jaime Lorenzo-Trueba, Gustav~Eje Henter, Shinji Takaki, Junichi Yamagishi,
  Yosuke Morino, and Yuta Ochiai,
\newblock ``Investigating different representations for modeling and
  controlling multiple emotions in dnn-based speech synthesis,''
\newblock {\em Speech Communication}, vol. 99, pp. 135--143, 2018.

\bibitem{r05}
Younggun Lee, Azam Rabiee, and Soo-Young Lee,
\newblock ``Emotional end-to-end neural speech synthesizer,''
\newblock {\em arXiv preprint arXiv:1711.05447}, 2017.

\bibitem{r06}
Yuxuan Wang, Daisy Stanton, Yu~Zhang, RJ-Skerry Ryan, Eric Battenberg, Joel
  Shor, Ying Xiao, Ye~Jia, Fei Ren, and Rif~A Saurous,
\newblock ``Style tokens: Unsupervised style modeling, control and transfer in
  end-to-end speech synthesis,''
\newblock in {\em International Conference on Machine Learning (ICML)}, 2018,
  pp. 5180--5189.

\bibitem{r07}
Ohsung Kwon, Inseon Jang, ChungHyun Ahn, and Hong-Goo Kang,
\newblock ``An effective style token weight control technique for end-to-end
  emotional speech synthesis,''
\newblock {\em IEEE Signal Processing Letters (SPL)}, vol. 26, no. 9, pp.
  1383--1387, 2019.

\bibitem{r08}
Se-Yun Um, Sangshin Oh, Kyungguen Byun, Inseon Jang, ChungHyun Ahn, and
  Hong-Goo Kang,
\newblock ``Emotional speech synthesis with rich and granularized control,''
\newblock in {\em IEEE International Conference on Acoustics, Speech and Signal
  Processing (ICASSP)}, 2020, pp. 7254--7258.

\bibitem{r09}
No{\'e} Tits, Kevin El~Haddad, and Thierry Dutoit,
\newblock ``Exploring transfer learning for low resource emotional tts,''
\newblock in {\em Proceedings of SAI Intelligent Systems Conference}, 2019, pp.
  52--60.

\bibitem{r10}
Pengfei Wu, Zhenhua Ling, Lijuan Liu, Yuan Jiang, Hongchuan Wu, and Lirong Dai,
\newblock ``End-to-end emotional speech synthesis using style tokens and
  semi-supervised training,''
\newblock in {\em Asia-Pacific Signal and Information Processing Association
  Annual Summit and Conference (APSIPA ASC)}, 2019, pp. 623--627.

\bibitem{r11}
Yang Gao, Weiyi Zheng, Zhaojun Yang, Thilo Kohler, Christian Fuegen, and Qing
  He,
\newblock ``Interactive text-to-speech via semi-supervised style transfer
  learning,''
\newblock {\em arXiv preprint arXiv:2002.06758}, 2020.

\bibitem{r12}
Carlos Busso, Murtaza Bulut, Chi-Chun Lee, Abe Kazemzadeh, Emily Mower, Samuel
  Kim, Jeannette~N Chang, Sungbok Lee, and Shrikanth~S Narayanan,
\newblock ``{IEMOCAP}: Interactive emotional dyadic motion capture database,''
\newblock {\em Language Resources and Evaluation}, vol. 42, no. 4, pp. 335,
  2008.

\bibitem{r13}
Fabien Ringeval, Andreas Sonderegger, Juergen Sauer, and Denis Lalanne,
\newblock ``Introducing the {RECOLA} multimodal corpus of remote collaborative
  and affective interactions,''
\newblock in {\em IEEE International Conference and Workshops on Automatic Face
  and Gesture Recognition (FG)}, 2013, pp. 1--8.

\bibitem{r14}
Carlos Busso, Srinivas Parthasarathy, Alec Burmania, Mohammed AbdelWahab,
  Najmeh Sadoughi, and Emily~Mower Provost,
\newblock ``Msp-improv: An acted corpus of dyadic interactions to study emotion
  perception,''
\newblock {\em IEEE Transactions on Affective Computing}, vol. 8, no. 1, pp.
  67--80, 2016.

\bibitem{r15}
Jianyou Wang, Michael Xue, Ryan Culhane, Enmao Diao, Jie Ding, and Vahid
  Tarokh,
\newblock ``Speech emotion recognition with dual-sequence lstm architecture,''
\newblock in {\em IEEE International Conference on Acoustics, Speech and Signal
  Processing (ICASSP)}, 2020, pp. 6474--6478.

\bibitem{r16}
Peng Song, Wenming Zheng, Shifeng Ou, Xinran Zhang, Yun Jin, Jinglei Liu, and
  Yanwei Yu,
\newblock ``Cross-corpus speech emotion recognition based on transfer
  non-negative matrix factorization,''
\newblock {\em Speech Communication}, vol. 83, pp. 34--41, 2016.

\bibitem{r17}
Yuan Zong, Wenming Zheng, Tong Zhang, and Xiaohua Huang,
\newblock ``Cross-corpus speech emotion recognition based on domain-adaptive
  least-squares regression,''
\newblock {\em IEEE signal processing letters}, vol. 23, no. 5, pp. 585--589,
  2016.

\bibitem{r19}
Siddique Latif, Junaid Qadir, and Muhammad Bilal,
\newblock ``Unsupervised adversarial domain adaptation for cross-lingual speech
  emotion recognition,''
\newblock in {\em International Conference on Affective Computing and
  Intelligent Interaction (ACII)}, 2019, pp. 732--737.

\bibitem{r20}
Zack Hodari, Oliver Watts, Srikanth Ronanki, and Simon King,
\newblock ``Learning interpretable control dimensions for speech synthesis by
  using external data.,''
\newblock in {\em Conference of the International Speech Communication
  Association (INTERSPEECH)}, 2018, pp. 32--36.

\bibitem{r21}
Arthur Gretton, Karsten~M Borgwardt, Malte~J Rasch, Bernhard Sch{\"o}lkopf, and
  Alexander Smola,
\newblock ``A kernel two-sample test,''
\newblock {\em The Journal of Machine Learning Research (JMLR)}, vol. 13, no.
  1, pp. 723--773, 2012.

\bibitem{r22}
Mingsheng Long, Yue Cao, Jianmin Wang, and Michael Jordan,
\newblock ``Learning transferable features with deep adaptation networks,''
\newblock in {\em International conference on machine learning (ICML)}, 2015,
  pp. 97--105.

\bibitem{r23}
Konstantinos Bousmalis, George Trigeorgis, Nathan Silberman, Dilip Krishnan,
  and Dumitru Erhan,
\newblock ``Domain separation networks,''
\newblock in {\em Advances in neural information processing systems}, 2016, pp.
  343--351.

\bibitem{r24}
Daniel Griffin and Jae Lim,
\newblock ``Signal estimation from modified short-time fourier transform,''
\newblock {\em IEEE Transactions on Acoustics, Speech, and Signal Processing},
  vol. 32, no. 2, pp. 236--243, 1984.

\bibitem{r25}
S.~King and Vasilis Karaiskos,
\newblock ``The blizzard challenge 2013,''
\newblock 2013.

\bibitem{r26}
Diederik~P. Kingma and Jimmy Ba,
\newblock ``Adam: {A} method for stochastic optimization,''
\newblock in {\em International Conference on Learning Representations (ICLR)},
  2015.

\bibitem{r28}
Laurens van~der Maaten and Geoffrey Hinton,
\newblock ``Visualizing data using t-sne,''
\newblock {\em Journal of machine learning research (JMLR)}, vol. 9, no. Nov,
  pp. 2579--2605, 2008.

\end{thebibliography}

\end{document}